# Study of thermodynamic properties of substitutional mixture of Co(II)- and Fe(II)-based octacyanoniobates

R. Pełka*,[1], P. Konieczny[1], Y. Miyazaki[2], Y. Nakazawa[2], T. Wasiutyński[1], A. Budziak[3], D. Pinkowicz[4], B. Sieklucka[4]

[1] *H. Niewodniczański Institute of Nuclear Physics, Polish Academy of Sciences, Radzikowskiego 152, 31-342 Kraków, Poland*
[2] *Research Center for Structural Thermodynamics, Graduate School of Science, Osaka University, Toyonaka, Osaka 560-0043, Japan*
[3] *Department of Hydrogen Energy, Faculty of Energy and Fuels, AGH University of Science and Technology, al. Mickiewicza 30, Kraków, Poland*
[4] *Faculty of Chemistry, Jagiellonian University, Ingardena 3, 30-060 Kraków, Poland*
* Corresponding author: *E-mail address:* Robert.Pelka@ifj.edu.pl

## Abstract

A comprehensive study of thermodynamic properties of three samples of bimetallic molecular magnets $[Co^{II}(pyrazole)_4]_{2x}[Fe^{II}(pyrazole)_4]_{2(1-x)}[Nb^{IV}(CN)_8]\cdot 4H_2O$ with $x=0$ (**Co₂Nb**), 0.5 (**CoFeNb**), and 1 (**Fe₂Nb**) is reported. The three samples display the same crystallographic structure crystallizing in the tetragonal system with space group $I4_1/a$. Their heat capacities are measured in the temperature range 0.36-100 K without applied field as well as in the field of $\mu_0 H$=0.1, 0.2, 0.5, 1, 2, 5, and 9 T. The results imply the presence of the second-order phase transitions to magnetically ordered phases at 4.87(8) K, 7.1(2) K, and 8.44(3) K for $x$=0, 0.5, and 1, respectively. The corresponding thermodynamic functions are analyzed to discuss the stability of the mixed compound and the magnetocaloric effect (MCE). The Gibbs energy of mixing is found to be positive but smaller in magnitude than the energy of thermal fluctuations indicating that the mixed sample is marginally stable in the full detected temperature range. The enthalpy of mixing is negative, which points to an ordered arrangement of the Co(II) and Fe(II) ions in the solid solution **CoFeNb**. The negative values of the entropy of mixing are explained by considering the enhanced rigidity of the crystal lattice of the solid solution sample. To extract the magnetic contribution to the heat capacity an approach based on a reasonable frequency spectrum is adopted. Taking advantage of the in-field heat capacity measurements MCE was described in terms of the isothermal entropy change $\Delta S_M$ and the adiabatic temperature change $\Delta T_{ad}$. The magnitudes of these quantities are typical for the class of molecular magnets. The values of $|\Delta S_M|^{max}$ detected for $\mu_0 \Delta H$=5 T amount to 7.04, 5.26, and 4.93 J K$^{-1}$ mol$^{-1}$ for **Co₂Nb**, **CoFeNb**, and **Fe₂Nb**, respectively, and are on the order of those obtained for the same field change in the isostructural compounds. The values of $\Delta T_{ad}$ detected for $\mu_0 \Delta H$=5 T amount to 4.16, 2.47, and 2.01 J K$^{-1}$ mol$^{-1}$ for **Co₂Nb**, **CoFeNb**, and **Fe₂Nb**, respectively, and are larger or comparable to those observed for the isostructural compounds. Temperature dependences of exponent $n$, quantifying the field dependence of $\Delta S_M$, display minima close to the transition temperatures implying through their values that the studied compounds belong to the universality class of the three-dimensional (3D) Heisenberg model. The regeneration Ericsson cycles employing the studied compounds as the working substance were considered. Most surprisingly, the Ericsson cycle operating between the temperatures corresponding to the full width at half-maximum of the

|Δ$S_M$| signal ($T_C$,$T_H$) turns out to be totally ineffective. Through shifting the temperature of the hot reservoir $T_H$ down to the temperature $T_{max}$ corresponding to |Δ$S_M$|$^{max}$ the coefficient of performance is rendered positive and comparable to that of the Carnot cycle. A detailed analysis indicates that the regeneration Ericsson cycle operating between $T_C$ and $T_{max}$ should be most efficient for the maximal studied value of the applied field (=9 T) with irrelevant differences between the studied compounds.

1. Introduction

For almost three decades now magnetic coordination networks have been given assiduous attention of both chemists and physicists, resulting in materials displaying a long-range magnetic order above room temperature [1-3]. Concerted efforts were focused on obtaining novel compounds displaying additional key features such as structural and electronic nonrigidity, noncentrosymmetry, chirality, host-guest behavior, luminescence, and others. They bore fruit in an array of novel classes of multifunctional magnetic materials with porous magnets, magnetic sponges, charge-transfer complexes, spin crossover magnets, photomagnets [4-13], noncentrosymmetric and chiral magnets [9, 10, 12, 14-18] luminescent magnets [19] or compounds exploiting the second-order magneto-optical or magneto-chiral effects [12, 14, 17, 18]. The quest for new coordination networks or clusters endowed with technologically relevant functionalities is by no means completed. In this respect, the magnetocaloric effect (MCE) [20] ranks among the emerging potential applications in cryogenic devices. Furthermore, one of the little exploited strategies to obtain novel molecular magnetic materials is the synthesis of substitutional mixtures displaying properties smoothly interpolating between those of the pure compounds [21-24]. This report goes along this late line of research.

Magnetocaloric effect, i.e. heating or cooling of a magnetic material following switching on or off of applied magnetic field in adiabatic conditions, is an area of intensive research due to the fact that magnetic refrigerators represent an environmentally friendly technology dispensing with media associated with ozone depletion or greenhouse effect. Moreover, they were demonstrated to show an enhanced efficiency in comparison to conventional gas compression-expansion refrigerators [25]. Crucial progress in this field involved on the one hand a considerable enhancement of material performance mostly realized through giant magnetocaloric effect (GMCE) [26, 27] and on the other hand the cost reduction through the replacement of rare earth elements by transition metal alloys [28]. However, in spite of the fact that GMCE occurring in materials with first-order magnetostructural phase transitions secures a large magnetic entropy change Δ$S_M$, it has two important deficiencies. The first is the narrowness of the Δ$S_M$ vs $T$ curve and the other - the presence of hysteresis leading to low operational frequencies and limited cooling power (refrigerant capacity RC). The profile of isothermal entropy change Δ$S_M$($T$) of materials undergoing the second-order phase transitions is by contrast more extended, albeit it is smaller in magnitude. What is more, thermal hysteresis is not exhibited by this class of materials. Therefore, a compromise between an optimal RC and the lack of thermal hysteresis makes them superior candidates for the development of magnetic cooling devices. Recent developments in MCE resulted in a new technological solution, where the magnetic entropy change $\Delta S_M$ is obtained by rotating a single crystal in a constant magnetic field [29-32]. This technique is confined to materials displaying considerable magnetic anisotropy. The rotating magnetocaloric effect (RMCE) introduces crucial improvement of magnetic cooling by obviating the need to move the single-crystal coolant in and out of a magnetic field or manipulate the field amplitude and replacing it by a much simply feasible change of the crystal orientation in a stationary field. Balli et al. [33, 34] devised a prototype of a rotary magnetic refrigerator, which integrates a simple construction with high efficiency, originating from the operation at higher frequencies than employed in conventional MCE, and a lower energy consumption due to the use of permanent magnets [33-35]. However, the latter feature puts a limit to the maximum field which can be employed for this type of cooling below 2 T. RMCE is a very young topic, which is expressed by exceptionally sparse subject literature concerning inorganic materials [29-34] as well as molecular compounds [36-39].

In the field of molecular magnetism MCE has been investigated most of all for single molecule magnets (SMMs) whose large ground-state spin value promises a substantial entropic effect [40-45]. Further studies were devoted to molecular rings, forming a subclass of molecular magnets characterized by a typical cyclic shape and a dominant antiferromagnetic coupling between the metallic nearest neighbors [46]. Besides, the very first studies of MCE associated with the second-order phase transition to a long-range magnetically ordered state involved Prussian blue analogs [47-49]. Further examples refer to a bimetallic octacyanoniobate $\{[M(II)(pyrazole)_4]_2[Nb(IV)(CN)_8]\cdot 4H_2O\}_n$ (M=Mn, Ni) isomorphous with the compounds under study [50, 51], an interesting instance of a molecular sponge changing reversibly the ordering temperature and the coercive field upon hydration/dehydration [52], and the effect of hydrostatic pressure on MCE in $Mn_2$-pyridazine-$[Nb(CN)_8]$ [53].

In this paper we are considering three samples of bimetallic molecular magnets with the formula $[Co^{II}(pyrazole)_4]_{2x}[Fe^{II}(pyrazole)_4]_{2(1-x)}[Nb^{IV}(CN)_8]\cdot 4H_2O$ (pyrazole is a five membered ring ligand $C_3H_4N_2$), where $x=1$ (**Co$_2$Nb**) [1], $x=0.5$ (**CoFeNb**), $x=0$ (**Fe$_2$Nb**) [54], where the middle compound represents a substitutional mixture of the two marginal ones referred to in what follows as the mixed and pure compounds, respectively. All three compounds are isostructural and crystallize in the tetragonal $I4_1/a$ space group. Their structure consists of a three-dimensional (3D) skeleton, where each Nb(IV) center is linked through the cyanido bridges M(II)-NC-Nb(IV) to four M(II) (M=Co, Fe) ions, whereas each M(II) center is bridged exclusively to two Nb(IV) ions. The remaining part of the distorted pseudooctahedral coordination sphere of M(II) is filled with pyrazole molecules, while the Nb(IV) ion coordinates further four terminal CN$^-$ ligands. The fact that such low connectivity indices should produce a 3D extended network represents the unique structural feature of these compounds. The graphical representation of the crystal structure of these compounds is shown in Fig. 1 for instant reference.

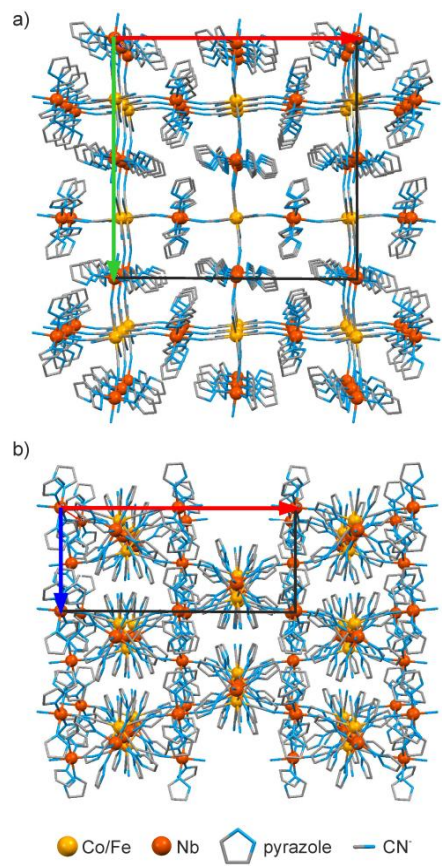

**Fig. 1:** Structure of $\{[M^{II}(pyrazole)_4]_2[Nb^{IV}(CN)_8]\cdot 4H_2O\}_n$. (a) View of the structure along the $c$ crystallographic axis (red and green mark the $a$ and $b$ crystallographic axis, respectively). The water molecules and hydrogen atoms are not shown for clarity. (b) View of the structure along the $b$ crystallographic axis (red and blue mark the $a$ and $c$ crystallographic axis, respectively).

The pure compounds **Co₂Nb** and **Fe₂Nb** were demonstrated to undergo a magnetic transition to a long-range order state at 5.9 K and 8.3 K, respectively, where the transition points were assumed to coincide with the position of the d$M$/d$T$ peak [54]. The analysis of the DC susceptibility and isothermal magnetization carried out in the framework of the molecular field model suggested that the character of the exchange coupling between the Co(II) and Nb(IV) centers is ferromagnetic, while that between the Fe(II) and Nb(IV) ions is antiferromagnetic with the exchange coupling constants estimated to amount to $J_{Co-Nb}$= +3.5(3) cm$^{-1}$, and $J_{Fe-Nb}$=-3.1(2) cm$^{-1}$ [54]. **Co₂Nb** was thus found to be a molecular ferromagnet, whereas **Fe₂Nb** - a molecular ferrimagnet. The critical behavior of **Fe₂Nb** studied with the use of ac magnetometry and zero-field μSR spectroscopy allowed to determine the static critical exponents $\beta$=0.42(3), $\gamma$=1.38(8), and the dynamic exponent $w$=0.33(2), thus placing the compound in the universality class of the 3D Heisenberg model [55]. Moreover, preliminary measurements of **Fe₂Nb** by relaxation calorimetry in zero and nonzero applied fields were analyzed to extract temperature dependences of the isothermal entropy change $\Delta S_M$ and the adiabatic temperature change $\Delta T_{ad}$ due to the magnetic field change, the two basic characteristics of the magnetocaloric effect (MCE). The maximum value of $\Delta S_M$ for $\mu_0\Delta H$=5 T was found to be placed at 10.3 K and amounts to 4.8 J mol$^{-1}$ K$^{-1}$, while the corresponding maximum value of $\Delta T_{ad}$=2.0 K was observed at 8.9 K [56]. The thermodynamic properties of the other pure compound **Co$_2$Nb** as well as the mixed one **CoFeNb** have not been reported previously.

In order to study thermodynamic properties of the reported compounds their heat capacities have been measured by relaxation calorimetry. The compounds are demonstrated to display transitions to magnetically long-range ordered states at 4.9 K (**Co$_2$Nb**), 7.3 K (**CoFeNb**), and 8.8 K (**Fe$_2$Nb**) as indicated by the positions at which the derivative d$C_p(T)$/d$T$ vanishes revealing the local maxima of $C_p(T)$. The measured values of $C_p$ are prerequisites of a detailed discussion of thermodynamic properties of the three samples with the focus on the mixed compound **CoFeNb**. The paper is organized as follows. On providing the experimental details in Section 2 we analyze and discuss the thermal behavior in zero applied field in Section 3. Then, in Section 4, we go on to describe an extraction procedure of the magnetic contribution to the heat capacity. Section 5 is devoted to the calculation and discussion of the magnetocaloric effect of the studied samples with some comments on its practical aspects. We close the paper in Section 6 with a bunch of general conclusions.

2. Experimental

The polycrystalline samples of **Co$_2$Nb**, **Fe$_2$Nb** were synthesized according to the procedures reported in [54], while **CoFeNb** solid state solution was synthesized according to a modified procedure. A solution of $(NH_4)_2Fe(SO_4)_2 \cdot 6H_2O$ (39 mg, 0.1 mmol), $CoCl_2 \cdot 6H_2O$ (23 mg, 0.1 mmol) and pyrazole (82 mg, 1.2 mmol) in degassed water (3 ml) was added dropwise to the degassed aqueous solution (3 ml) of $K_4Nb(CN)_8 \cdot 2H_2O$ (50 mg, 0.1 mmol). A violet precipitate formed immediately. The suspension was stirred for 5 min, filtered and dried shortly in air. The violet powder was stored at low temperature due to slight sensitivity to air. Anal. Calcd for $C_{32}CoFeH_{40}N_{24}NbO_4$: C, 37.22; H, 3.90; N, 32.56. Found: C,37.11; H, 3.72; N, 32.73. The powder XRD pattern for **CoFeNb** was recorded at room temperature using a PANalytical X'Pert PRO MPD diffractometer (Cu K$\alpha$ X-ray radiation 1.541874 Å). It was identified with the FullProf program [57] based on the Rietveld method [58]. The best fit was obtained with the assumption of a tetragonal structure identical to that for **Co$_2$Nb**, **Fe$_2$Nb** (space group: I4$_1$/a; lattice parameters: $a$ = 21.633(6) Å, $c$ = 9.619(4) Å; cell volume: 4501.6 Å$^3$; R-factors, not corrected for background: Rp=4.79, Rwp=6.61, Rexp=4.70, $\chi^2$=1.97), see Fig. 2. There are no traces of distortion or the second phase. For the **CoFeNb** compound, it is not possible to distinguish the positions of Co and Fe atoms in the PXRD pattern as they occupy the same 8$c$ Wyckoff position.

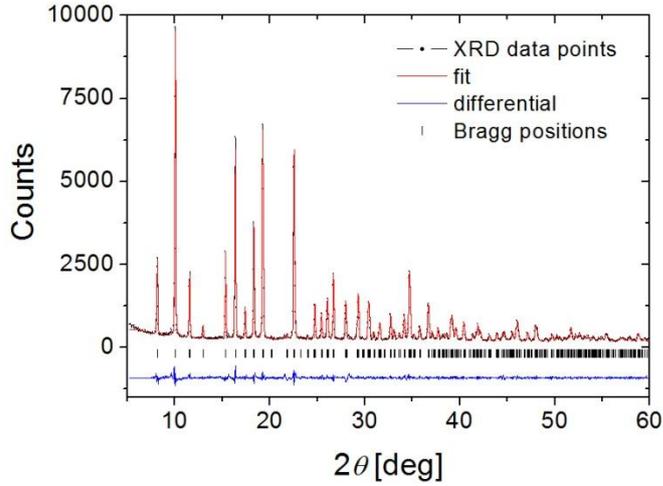

**Fig. 2:** Observed (symbols), calculated (solid line), and difference (observed-calculated; bottom solid line) X-ray diffraction profiles at 300 K for **CoFeNb** indicate the positions of the Bragg reflections for tetragonal phase with space group $I4_1/a$.

The heat capacity measurements were carried out with the PPMS Quantum Design instrument by the relaxation calorimetry technique. The measurements were performed in two independent stages. In the high-temperature stage, using polycrystalline samples of mass 1.7504 mg (**Co$_2$Nb**), 2.2627 mg (**CoFeNb**), and 2.0087 mg (**Fe$_2$Nb**) pressed to form small pellets, the heat capacity was detected in the cooling direction in the temperature range of 1.9–101 K without applied field by the standard probe cooled with $^4$He. In the low-temperature stage, employing the $^3$He probe and using polycrystalline samples of mass 1.5417 mg (**Co$_2$Nb**), 2.4641 mg (**CoFeNb**), and 3.3189 mg (**Fe$_2$Nb**) pressed to form small pellets, the heat capacity measurements were run in the cooling direction in the temperature range of 0.36–20.2 K without applied field as well as in the field of $\mu_0H$=0.1, 0.2, 0.5, 1, 2, 5, and 9 T. Since the heat capacity data measured with the $^3$He system at high temperatures and the data detected with the $^4$He system at low temperatures are generally incorrect, we decided to use the zero-field data provided by the former system below 20 K and those provided by the latter system above 20 K.

3. Zero-field heat capacity

Just in the beginning let us focus on the zero-field data. Figure 3 shows the zero-field heat capacity in the range 0.36 - 20 K for the studied samples. It can be seen that all three samples reveal anomalies which can be assigned to the second-order phase transition to a magnetically ordered phase. The transition temperatures of the compounds are 4.9 K (**Co$_2$Nb**), 7.3 K (**CoFeNb**), and 8.8 K (**Fe$_2$Nb**) as indicated by the positions at which the derivative $dC_p(T)/dT$ vanishes revealing the local maxima of $C_p(T)$. Let us stress that for the mixed compound **CoFeNb** we observe a single anomaly with the transition temperature placed between the transition temperatures of the pure samples **Co$_2$Nb** and **Fe$_2$Nb**, which implies that the sample is a single-phase one with the metal sites occupied randomly by the Co(II) and Fe(II) ions according to the expected ratio 1:1 (a proper mixture) and thus no phase decomposition takes place. Moreover, the transition temperature of **CoFeNb** compares well with the value of 7.1 K predicted by the molecular field model, see Eq. (A7) in Appendix.

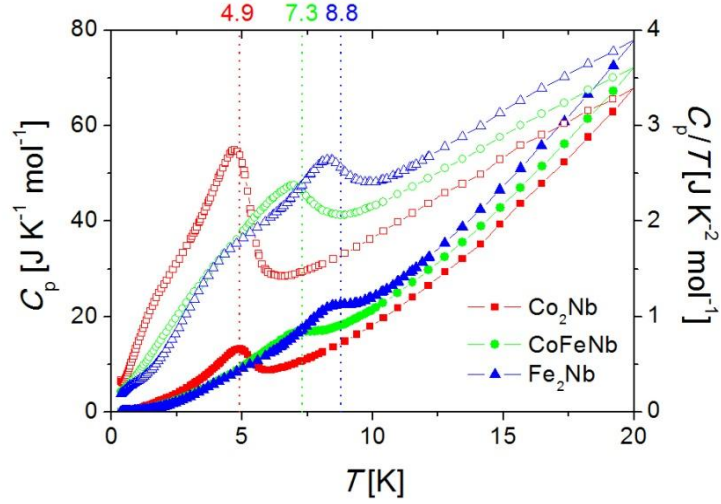

**Fig. 3:** Zero-field heat capacity in terms of $C_p$ (full symbols) and $C_p/T$ (open symbols) of the studied samples in the low-temperature regime.

The following analysis aims to calculate consistently the thermodynamic functions on the basis of the heat capacity data. To do it correctly one needs the extrapolation of the heat capacity data in the experimentally missing temperature interval of 0-0.36 K. The correct extrapolation must take into account the nuclear contributions to the heat capacity. Table 1 shows the spin states together with the abundances of the constituent nuclei.

**Table 1:** Constituent metal nuclei [59].

| Nucleus | spin quantum number $I$ | abundance [%] |
|---|---|---|
| $^{59}$Co | 7/2 | 100 |
| $^{54}$Fe | 0 | 5.845 |
| $^{56}$Fe | 0 | 91.754 |
| $^{57}$Fe | 1/2 | 2.119 |
| $^{58}$Fe | 0 | 0.282 |
| $^{93}$Nb | 9/2 | 100 |

It thus turns out that we have to account for the non-zero nuclear spin of the Co and Nb ions, while the nuclear contribution of the Fe ion may be safely neglected. To quantify the nuclear contribution to the heat capacity we assume that the ground state of either nucleus is split due to the local magnetic field. The energy of this splitting is given by the formula

$$E(M_I) = AM_I \quad (M_I = -I, -I+1, \ldots, I-1, I),$$

where $A$ (in units of energy) is proportional to the local field strength, and $I$ denotes the nuclear spin quantum number. The corresponding contribution to the molar specific heat has the following form

$$C_n(\beta; A, I) = \frac{1}{4}R(\beta A)^2 \left\{ 4I(I+1) - (2I+1)^2 \operatorname{ctgh}^2\left[\frac{2I+1}{2}\beta A\right] + \operatorname{ctgh}^2\left[\frac{\beta A}{2}\right] \right\}, \quad (1)$$

where $\beta = 1/k_B T$. Apart from the contribution from the nuclear magnetism, at low temperatures we expect a contribution from the ionic magnetism and lattice vibrations. We represent these contribution by a single algebraic term $BT^C$. Thus the low temperature molar specific heat of the samples takes on the form

$$C_{LT}(Co_2Nb) = 2C_n(\beta; A_{Co}, 7/2) + C_n(\beta; A_{Nb}, 9/2) + B_1 T^{C_1} \quad (2)$$

$$C_{LT}(CoFeNb) = C_n(\beta; A_{Co}, 7/2) + C_n(\beta; A_{Nb}, 9/2) + B_2 T^{C_2} \quad (3)$$

$$C_{LT}(Fe_2Nb) = C_n(\beta; A_{Nb}, 9/2) + B_3 T^{C_3} \quad (4)$$

We fitted formula $C_{LT}(Co_2Nb)$ to the experimental data within the temperature range 0.36-0.75 K and obtained a plausible set of best-fit parameters: $A_{Co}=0.0076(5)$ K, $A_{Nb}=0.0061(3)$ K, $B_1=0.746(9)$ J K$^{-1}$ mol$^{-1}$, $C_1=2.47(7)$. Next we performed fitting of formulas $C_{LT}(CoFeNb)$ and $C_{LT}(Fe_2Nb)$ to the experimental data in the same temperature range keeping $A_{Co}$ and $A_{Nb}$ fixed at the above values and relaxing only parameters $B_i$ and $C_i$ ($i=2,3$), which yielded $B_2=0.44(1)$ J K$^{-1}$ mol$^{-1}$, $C_2=2.19(5)$, $B_3=0.368(8)$ J K$^{-1}$ mol$^{-1}$, $C_3=1.85(4)$. The values of exponents $C_i$ ($i=1,2,3$) are placed between 1.5 and 3, which might be expected for the temperature interval below 1 K, where the magnetic contribution ($\sim T^{3/2}$) dominates the lattice contribution ($\sim T^3$). Furthermore, using the fitted values of $A_{Co}$ and $A_{Nb}$ and employing the relation

$$B_{int} = \frac{k_B A[K] I}{\mu[\mu_N] \mu_N [J/T]}, \quad (5)$$

where $\mu_N = 5.050783699(31) \times 10^{-27}$ J T$^{-1}$ is the nuclear magneton [60], and $\mu$ is the nuclear moment expressed in nuclear magnetons ($\mu_{Co}=4.627(9)$, $\mu_{Nb}=6.1705(3)$ [61]), one can estimate the magnitude of the local magnetic field. One hence obtains $B_{int,Co}=15.8$ T at the Co nucleus and $B_{int,Nb}=12.2$ T at the Nb nucleus, which both are on the plausible order of magnitude. So it may be concluded that the extrapolation of the heat capacities down to 0 K is credible. Fig. 4 shows the result of the extrapolation. The Schottky anomalies corresponding to the nuclear magnetism of the samples are apparent.

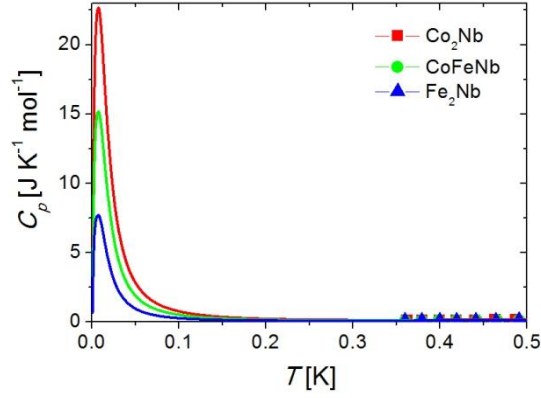

**Fig. 4:** Extrapolation of the zero-field heat capacity data down to 0 K. The Schottky anomalies corresponding to the nuclear magnetism of the samples are apparent.

Now that the extrapolations are ready to use we can go over to determining the thermodynamic functions. They were numerically calculated using the formulas

$$H(T) = \int_0^T C_p(T')\,dT' \tag{6}$$

$$S(T) = \int_0^T \frac{C_p(T')}{T'}\,dT' \tag{7}$$

$$G(T) = H(T) - TS(T) \tag{8}$$

Figures 5, 6, and 7 show the temperature dependence of the enthalpies, entropies, and the Gibbs free energies of the studied compounds, respectively

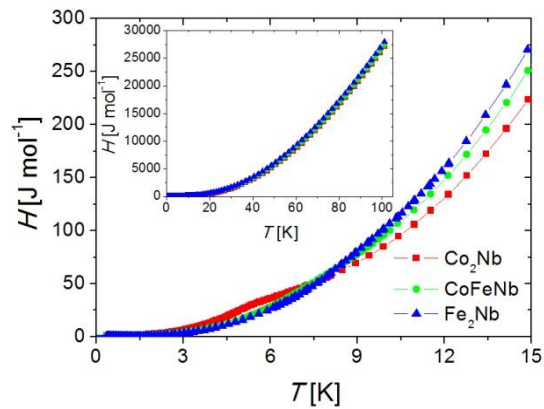

**Fig. 5:** Temperature dependence of the enthalpies of the studied samples. Inset: The full range data.

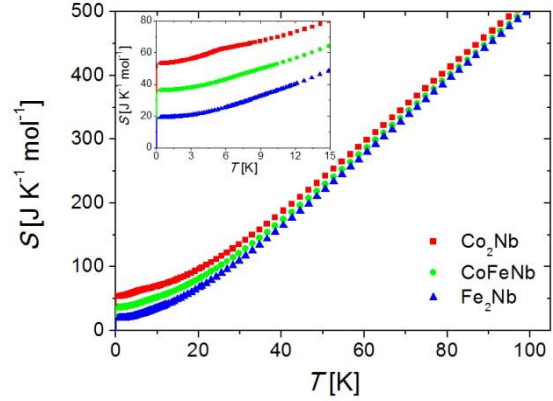

**Fig. 6:** Temperature dependence of the entropies of the studied samples. Inset: The low-temperature close-up.

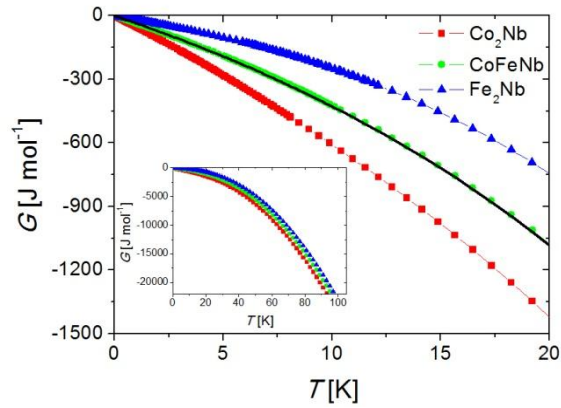

**Fig. 7:** Temperature dependence of the Gibbs free energies of the studied samples. Inset: The full range data.

Figure 7 shows the close-up of the Gibbs free energies of the studied samples. The points corresponding to the mixed sample **CoFeNb** are placed in between the points corresponding to the pure compounds **Co₂Nb** and **Fe₂Nb**, which might have been expected. The black solid line shows the value of the equally weighted average of the Gibbs energies of the pure compounds: $G_{avg}=[G(\mathbf{Co_2Nb})+G(\mathbf{Fe_2Nb})]/2$. The points of the mixed sample (green circles) lie slightly above the black curve, indicating that there is a positive excess Gibbs energy. Figure 8 shows the temperature dependence of the excess Gibbs energy $\Delta G_{mix}=G(\mathbf{CoFeNb})-G_{avg}$, the excess enthalpy $\Delta H_{mix}=H(\mathbf{CoFeNb})-H_{avg}$, and the excess entropy $\Delta S_{mix}=S(\mathbf{CoFeNb})-S_{avg}$ which can be interpreted as the Gibbs free energy, the enthalpy and the entropy of mixing, respectively.

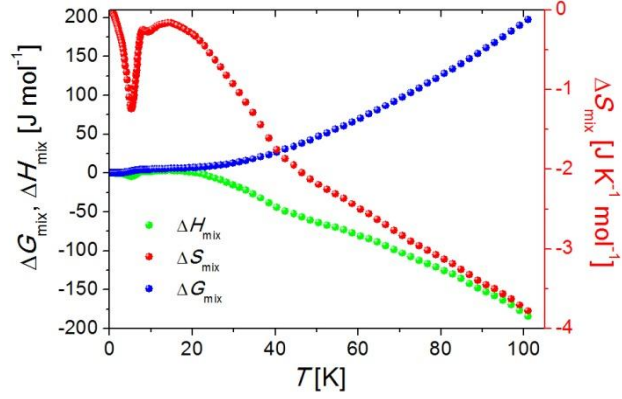

**Fig. 8:** Temperature dependence of the excess Gibbs free energy, the excess enthalpy and the excess entropy of the solid state solution **CoFeNb**.

Indeed, the Gibbs free energy of mixing $\Delta G_{mix}$ is positive in the full temperature range, which suggests that the mixed sample **CoFeNb** is unstable. However, for the sake of argument, let us express the value of $\Delta G_{mix}$ observed at 100 K in terms of temperature, i.e. $\Delta T_{mix}=\Delta G_{mix}/R$, where $R=8.3144598(48)$ J K$^{-1}$ mol$^{-1}$ is the molar gas constant [60]. One readily obtains that $\Delta T_{mix}=23.7$ K, which is one order of magnitude lower than the average energy of thermal fluctuations at that temperature (100 K). Thus, the Gibbs energy of mixing is admittedly positive, however it is smaller in magnitude than the energy of thermal fluctuations, which renders the sample marginally stable in the full detected temperature range.

At the same time, the enthalpy of mixing $\Delta H_{mix}$ is negative in the full temperature range. To understand the implications of this fact let us assume that there are $N$ lattice sites in the crystal structure on which substitution takes place and each such lattice site is surrounded by $z$ nearest neighboring sites belonging to this set. Then the total number of the nearest-neighbor bonds is $1/2Nz$, where the factor $1/2$ arises since there are two ions per bond. Next, let the energy associated with Co-Co, Fe-Fe and Co-Fe nearest neighbor pairs be, respectively, $E_{CoCo}$, $E_{FeFe}$ and $E_{CoFe}$. If the Co and Fe ions are mixed randomly, then the probability of Co-Co, Fe-Fe and Co-Fe neighbors is $x^2$, $(1-x)^2$ and $2x(1-x)$, respectively, where $x$ denotes the mole fraction of the Co ions. Hence the total enthalpy of the solid solution is given by

$$H = \frac{1}{2}Nz\left[x^2 E_{CoCo} + (1-x)^2 E_{FeFe} + 2x(1-x) E_{CoFe}\right] \quad (9)$$

which can be rearrange to read

$$H = \frac{1}{2}Nz[x E_{CoCo} + (1-x) E_{FeFe}] + \frac{1}{2}Nzx(1-x)(2E_{CoFe} - E_{CoCo} - E_{FeFe}). \quad (10)$$

The first term in Eq. (10) corresponds to the enthalpy of the mechanical mixture, while the second term may be interpreted as the excess enthalpy of mixing $\Delta H_{mix}$. Its sign is determined by the sign of the interaction parameter $E_{int}=2E_{CoFe}-E_{CoCo}-E_{FeFe}$. The negative value of $E_{int}$ ($E_{CoFe}<(E_{CoCo}+E_{FeFe})/2$), which is the case with **CoFeNb** ($\Delta H_{mix}<0$), indicates that it is energetically more favorable to have Co-Fe neighbors, rather than Co-Co or Fe-Fe neighbors.

To minimize the internal energy of the compound the number of Co-Fe neighbors should be maximized, thus the solid solution is expected to form an ordered compound with Co and Fe ions filling the sites in an alternating fashion.

Although the negative values of the enthalpy of mixing $\Delta H_{mix}$ can be understood in terms of an ordered arrangement of the Co and Fe ions in the solid solution **CoFeNb**, the fact that the excess entropy $\Delta S_{mix}$ is also negative in the full temperature range is more surprising and more challenging to rationalize. Assuming that the entropy of the solid solution is mainly vibrational in origin, i.e. related to the structural disorder caused by thermal vibrations of the atoms at finite temperature, let us consider a toy model of a one-dimensional lattice whose sites may be occupied by two types of atoms A and B. The pure compounds **Co₂Nb** and **Fe₂Nb** correspond to the configurations where the sites are occupied exclusively either by A-type atoms or by B-type atoms, while the solid solution **CoFeNb** is tantamount to the configuration where there are exactly as many A-type atoms as the B-type atoms in the system and either they are randomly distributed over the lattice sites or their distribution is ordered with an alternating arrangement of the atoms forming the ...ABABAB... chain. Figure 9 shows the four different configurations.

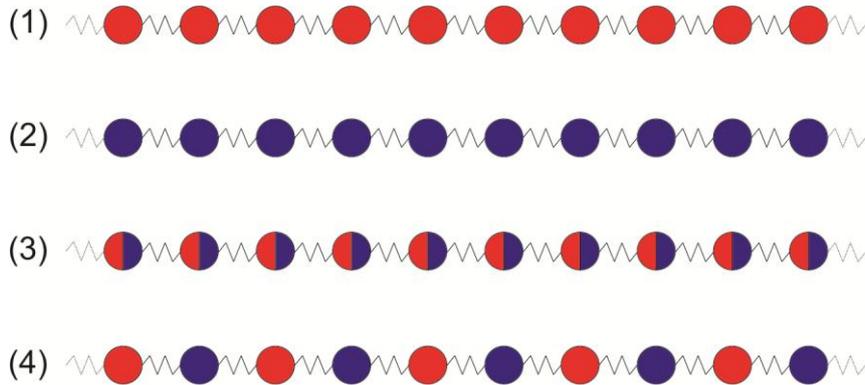

**Fig. 9:** The four different configurations of the one-dimensional lattice considered in the toy model. (1) The configuration corresponding to the pure compound **Co₂Nb.** (2) The configuration corresponding to the pure compound **Fe₂Nb**. (3) The configuration corresponding to the solid solution **CoFeNb** with the disordered distribution of the constituent atoms. (4) The configuration corresponding to the solid solution **CoFeNb** with the ordered distribution of the constituent atoms.

For configurations (1)-(3) the lattice involves a monoatomic basis with atoms of mass $m_1$, $m_2$, and $(m_1+m_2)/2$, respectively. The corresponding phonon spectrum is defined by the dispersion relation

$$\omega^{(i)}(ka) = 2\sqrt{\frac{k_f}{m^{(i)}}}\left|\sin\left(\frac{1}{2}ka\right)\right| \quad \text{with} \quad -\pi \leq ka \leq \pi \quad (i=1,2,3), \qquad (11)$$

where $k_f$ is the force constant, $m^{(i)}$ is the mass of the atom, $k$ is the wavevector, and $a$ - the distance between the nearest neighbors (the lattice constant). The basis of the lattice configuration (4) is diatomic (there is an alternating arrangement of atoms of mass $m_1$ and

$m_2$), which leads to the presence of two branches in the phonon spectrum with the following dispersion relation

$$\omega_\pm^{(4)}(ka) = \sqrt{\frac{k_f}{\mu}} \sqrt{1 \pm \sqrt{1 - 4\left(\frac{\mu}{m}\right)^2 \sin^2(ka)}} \quad \text{with} \quad -\frac{\pi}{2} \leq ka \leq \frac{\pi}{2}, \tag{12}$$

where "+" corresponds to the optical branch, while "−" - to the acoustic branch, $\mu = m_1 m_2/(m_1+m_2)$, $m = \sqrt{m_1 m_2}$, and $a$ is still the distance between the nearest neighbors (the lattice constant being now $2a$). Knowing that the entropy associated with a single phonon mode of circular frequency $\omega$ at temperature $T$ is given by the formula

$$S_{\text{vib}}(\omega)/k_B = \frac{\beta\hbar\omega}{e^{\beta\hbar\omega} - 1} - \ln(1 - e^{-\beta\hbar\omega}), \tag{13}$$

where $\beta=1/k_B T$, one can calculate the molar vibrational entropy of the four lattice configurations in the continuous limit using the formulas

$$S_{\text{vib}}^{(i)}/R = \frac{3}{2\pi} \int_{-\pi}^{\pi} d\xi \, S_{\text{vib}}[\omega^{(i)}(\xi)] \quad (i=1,2,3), \tag{14}$$

$$S_{\text{vib}}^{(4)}/R = \frac{3}{2\pi} \int_{-\pi/2}^{\pi/2} d\xi \left( S_{\text{vib}}[\omega_+(\xi)] + S_{\text{vib}}[\omega_-(\xi)] \right), \tag{15}$$

where $R$ is the molar gas constant, and factor "3" accounts for one longitudinal and two transverse modes. Assuming $m_1 = m_{\text{Co}} = 58.933194$ u, $m_2 = m_{\text{Fe}} = 55.845$ u [60], $k_f = 520$ N m$^{-1}$ (typical for single bond like in HCl), which returns the reference energy scale of lattice vibrations $T_{\text{vib},0} = \hbar\sqrt{k_f/\mu}/k_B = 800$ K, and allowing for a change of $T_{\text{vib}} = T_{\text{vib},0}(1+\delta[\%]/100)$ within some range for mixed configurations (3) and (4) as induced by a possible change of the force constant $k_f$, we calculated the corresponding entropies of mixing defined by

$$\Delta S_{\text{mix}}^{(j)}(\delta) = S_{\text{vib}}^{(j)}(\delta) - \frac{S_{\text{vib}}^{(1)} + S_{\text{vib}}^{(2)}}{2} \quad (j=3,4). \tag{16}$$

Figure 10 shows the temperature dependence of $\Delta S_{\text{mix}}^{(4)}(\delta)$ for an array of values of $\delta$=-10, -5, -3, -1, 0, 1, 3, 5, 10 % (a similar plot for $\Delta S_{\text{mix}}^{(3)}(\delta)$ is not shown for being almost the same). It can be seen that stiffening of the lattice ($\delta>0$), which may be due to the requirement of accommodating two different atoms within the same lattice structure, leads to negative values of the entropy of mixing. This is most probably the case with our solid solution sample **CoFeNb**.

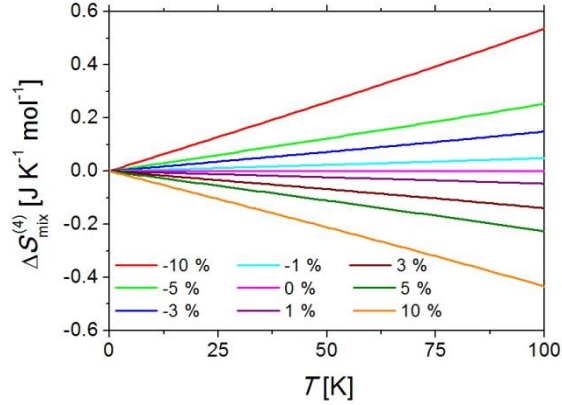

**Fig. 10:** The temperature dependence of $\Delta S_{\text{mix}}^{(4)}(\delta)$ for several indicated values of $\delta$. ). It is apparent that stiffening of the lattice ($\delta>0$) leads to negative values of the entropy of mixing.

Finally, we want to analyze the entropy content of the studied samples associated with the discrete degrees of freedom, i.e. nuclear and ionic spins. For this purpose, we choose to look at the entropy differences between the samples assuming that the lattice contributions in all three cases are comparable and will cancel each other out if we take duly into account the entropy of mixing for **CoFeNb** which was shown above to be vibrational in origin. Figure 11 shows the result of the subtractions.

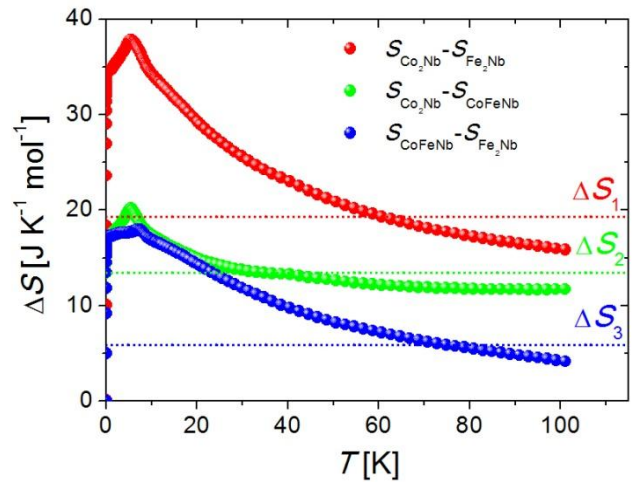

**Fig. 11:** Temperature dependence of the entropy differences between the studied samples.

All three differences indicated in Fig. 11 were constructed so as to reveal positive values in the whole temperature range. The differences are peaked at low temperatures, which is associated with the entropy contributions originating from the nuclear spins $S_n(\text{Co})=R\ln(2\times 7/2+1)\approx 17.29$ J K$^{-1}$ mol$^{-1}$ and $S_n(\text{Nb})=R\ln(2\times 9/2+1)\approx 19.14$ J K$^{-1}$ mol$^{-1}$: with $S_n(\textbf{Co}_2\textbf{Nb})\approx 2S_n(\text{Co})+S_n(\text{Nb})$, $S_n(\textbf{Fe}_2\textbf{Nb})\approx S_n(\text{Nb})$, and $S_n(\textbf{CoFeNb})\approx S_n(\text{Co})+S_n(\text{Nb})$ one arrives at $S_n(\textbf{Co}_2\textbf{Nb})-S_n(\textbf{Fe}_2\textbf{Nb})\approx 2S_n(\text{Co})\approx 34.58$ J K$^{-1}$ mol$^{-1}$, $S_n(\textbf{Co}_2\textbf{Nb})-S_n(\textbf{CoFeNb})\approx S_n(\text{Co})\approx 17.29$ J K$^{-1}$ mol$^{-1}$, and $S_n(\textbf{CoFeNb})-S_n(\textbf{Fe}_2\textbf{Nb})\approx S_n(\text{Co})\approx 17.29$ J K$^{-1}$ mol$^{-1}$, which is roughly consistent with the peak heights. With increasing temperature the differences decrease and approach something close to saturation at the high temperature limit, which should be so if the lattice contributions should actually cancel out. We can estimate the values of the entropy differences expected at high temperatures by assuming that the ionic spin contributions are saturated there. The spin of the Fe(II) ion is $S_{Fe}=2$, the spin of the Nb(IV) ion amounts to $S_{Nb}=1/2$, while the spin of the Co(II) ion is taken to be equal to $S_{Co}=1/2$, as we assume that below 100 K essentially only the ground sate of the Co(II) ion is populated which is the Kramers doublet originating from the crystal-field split multiplet corresponding to the full spin of 3/2. One thus obtains

$$S_{Co_2Nb} = R[2\ln(2S_{Co}+1) + \ln(2S_{Nb}+1)] + 2S_n(Co) + S_n(Nb) \approx 71.01 \text{ J K}^{-1} \text{ mol}^{-1} \quad (17)$$

$$S_{CoFeNb} = R[\ln(2S_{Co}+1) + \ln(2S_{Fe}+1) + \ln(2S_{Nb}+1)] + S_n(Co) + S_n(Nb) + \delta S \approx 57.56 \text{ J K}^{-1} \text{ mol}^{-1} \quad (18)$$

$$S_{Fe_2Nb} = R[2\ln(2S_{Fe}+1) + \ln(2S_{Nb}+1)] + S_n(Nb) \approx 51.67 \text{ J K}^{-1} \text{ mol}^{-1} \quad (19)$$

where $\delta S = \Delta S_{mix}(100K) \approx -3.78$ J K$^{-1}$ mol$^{-1}$ is a correction accounting for the changes of the lattice contribution of the mixed compound. The corresponding estimates read

$$\Delta S_1 = S_{Co_2Nb} - S_{Fe_2Nb} \approx 19.34 \text{ J K}^{-1} \text{ mol}^{-1} \quad (20)$$

$$\Delta S_2 = S_{Co_2Nb} - S_{CoFeNb} \approx 13.45 \text{ J K}^{-1} \text{ mol}^{-1} \quad (21)$$

$$\Delta S_3 = S_{CoFeNb} - S_{Fe_2Nb} \approx 5.89 \text{ J K}^{-1} \text{ mol}^{-1} \quad (22)$$

and are depicted in Fig. 11 with dotted lines. It can be seen that they roughly agree with the observed values corroborating the assumed ionic spin content of the samples.

4. Magnetic contribution to heat capacity

In general, the measured heat capacity is a conglomerate of several contributions stemming from various degrees of freedom. Except for the fortuitous case of a sharp transition phenomenon, the task to subdivide a total heat capacity among each contribution is far from straightforward. One of the most widely adopted approaches is to calculate the lattice heat capacity by adopting a plausible Debye temperature or a pair of Debye and/or Einstein temperatures [62-64]. A corresponding states method [65], a temperature derivative method [66] or a temperature-dependent Debye temperature method [67] count among other successful case-sensitive methods of separation. In the case under study, however, all these methods cannot come in useful as the heat capacity anomalies due to magnetic interaction are extended over a wide temperature region. We therefore take an approach whose main principle is to calculate the lattice contribution based on a reasonable frequency spectrum, see [Sorai2] for detail. The measured heat capacities in the full temperature range were fitted using the function

$$C_p = C_{\text{normal}} + cT^{-2} ,  \quad (23)$$

where

$$C_{\text{normal}} = \text{Debye functions} + \text{Einstein functions} \quad (24)$$

is the sought for lattice contribution to the heat capacity. The characteristic frequencies of the Einstein contributions were based on the known intramolecular vibrations for pyrazole [68], the water molecule [69] and the CN⁻ ligand [54], while the Debye cutoff frequencies were determined through the fit. The second term in Eq. (23) corresponds to the high temperature contribution due to the magnetic short-range order. Table 2 lists the best fit parameters for the three studied compounds.

**Table 2: The best fit parameters yielded in the procedure determining the normal heat capacities**

| Co₂Nb | | | CoFeNb | | | Fe₂Nb | | |
|---|---|---|---|---|---|---|---|---|
| $c$ [J K mol⁻¹] | 1506.21 | | $c$ [J K mol⁻¹] | 3912.30 | | $c$ [J K mol⁻¹] | 4534.10 | |
| **known Einstein functions (> 700 cm⁻¹)** | | | | | | | | |
| symbol | frequency [cm⁻¹] | degeneracy | symbol | frequency [cm⁻¹] | degeneracy | symbol | frequency [cm⁻¹] | degeneracy |
| $\nu_{E1}$ | 3755.79 | 4 | $\nu_{E1}$ | 3755.79 | 4 | $\nu_{E1}$ | 3755.79 | 4 |
| $\nu_{E2}$ | 3656.65 | 4 | $\nu_{E2}$ | 3656.65 | 4 | $\nu_{E2}$ | 3656.65 | 4 |
| $\nu_{E3}$ | 3523.2 | 8 | $\nu_{E3}$ | 3523.2 | 8 | $\nu_{E3}$ | 3523.2 | 8 |
| $\nu_{E4}$ | 3154.5 | 8 | $\nu_{E4}$ | 3154.5 | 8 | $\nu_{E4}$ | 3154.5 | 8 |
| $\nu_{E5}$ | 3136.5 | 8 | $\nu_{E5}$ | 3136.5 | 8 | $\nu_{E5}$ | 3136.5 | 8 |
| $\nu_{E6}$ | 3125.8 | 8 | $\nu_{E6}$ | 3125.8 | 8 | $\nu_{E6}$ | 3125.8 | 8 |
| $\nu_{E7}$ | 2123 | 8 | $\nu_{E7}$ | 2123 | 4 | $\nu_{E7}$ | 2119 | 8 |
| $\nu_{E8}$ | 1594.59 | 4 | $\nu_{E8}$ | 2119 | 4 | $\nu_{E8}$ | 1594.59 | 4 |
| $\nu_{E9}$ | 1530.9 | 8 | $\nu_{E9}$ | 1594.59 | 4 | $\nu_{E9}$ | 1530.9 | 8 |
| $\nu_{E10}$ | 1447.2 | 8 | $\nu_{E10}$ | 1530.9 | 8 | $\nu_{E10}$ | 1447.2 | 8 |
| $\nu_{E11}$ | 1394.5 | 8 | $\nu_{E11}$ | 1447.2 | 8 | $\nu_{E11}$ | 1394.5 | 8 |
| $\nu_{E12}$ | 1357.5 | 8 | $\nu_{E12}$ | 1394.5 | 8 | $\nu_{E12}$ | 1357.5 | 8 |
| $\nu_{E13}$ | 1254.0 | 8 | $\nu_{E13}$ | 1357.5 | 8 | $\nu_{E13}$ | 1254.0 | 8 |
| $\nu_{E14}$ | 1158.6 | 8 | $\nu_{E14}$ | 1254.0 | 8 | $\nu_{E14}$ | 1158.6 | 8 |
| $\nu_{E15}$ | 1121.0 | 8 | $\nu_{E15}$ | 1158.6 | 8 | $\nu_{E15}$ | 1121.0 | 8 |
| $\nu_{E16}$ | 1054.3 | 8 | $\nu_{E16}$ | 1121.0 | 8 | $\nu_{E16}$ | 1054.3 | 8 |
| $\nu_{E17}$ | 1008.6 | 8 | $\nu_{E17}$ | 1054.3 | 8 | $\nu_{E17}$ | 1008.6 | 8 |
| $\nu_{E18}$ | 923.6 | 8 | $\nu_{E18}$ | 1008.6 | 8 | $\nu_{E18}$ | 923.6 | 8 |
| $\nu_{E19}$ | 908.2 | 8 | $\nu_{E19}$ | 923.6 | 8 | $\nu_{E19}$ | 908.2 | 8 |
| $\nu_{E20}$ | 878.8 | 8 | $\nu_{E20}$ | 908.2 | 8 | $\nu_{E20}$ | 878.8 | 8 |
| $\nu_{E21}$ | 832.9 | 8 | $\nu_{E21}$ | 878.8 | 8 | $\nu_{E21}$ | 832.9 | 8 |
| $\nu_{E22}$ | 745.0 | 8 | $\nu_{E22}$ | 832.9 | 8 | $\nu_{E22}$ | 745.0 | 8 |
| | | | $\nu_{E23}$ | 745.0 | 8 | | | |
| **Debye functions** | | | | | | | | |
| symbol | frequency [cm⁻¹] | degeneracy | symbol | frequency [cm⁻¹] | degeneracy | symbol | frequency [cm⁻¹] | degeneracy |
| $\nu_{D1}$ | 70 | 14.08 | $\nu_{D1}$ | 80 | 19.86 | $\nu_{D1}$ | 80 | 21.90 |
| $\nu_{D2}$ | 140 | 34.95 | $\nu_{D2}$ | 170 | 42.57 | $\nu_{D2}$ | 170 | 39.93 |
| **box Einstein functions (<700 cm⁻¹)** | | | | | | | | |

| symbol | frequency [cm$^{-1}$] | degeneracy | symbol | frequency [cm$^{-1}$] | degeneracy | symbol | frequency [cm$^{-1}$] | degeneracy |
|---|---|---|---|---|---|---|---|---|
| $\nu_{E1,L}$ | 140 | 48.91 | $\nu_{E1,L}$ | 170 | 33.89 | $\nu_{E1,L}$ | 170 | 37.17 |
| $\nu_{E1,H}$ | 326.67 | 48.91 | $\nu_{E1,H}$ | 346.67 | 33.89 | $\nu_{E1,H}$ | 346.67 | 37.17 |
| $\nu_{E2,L}$ | 326.67 | 13.44 | $\nu_{E2,L}$ | 346.67 | 15.06 | $\nu_{E2,L}$ | 346.67 | 3.67 |
| $\nu_{E2,H}$ | 513.33 | 13.44 | $\nu_{E2,H}$ | 523.33 | 15.06 | $\nu_{E2,H}$ | 523.33 | 3.67 |
| $\nu_{E3,L}$ | 513.33 | 7.39 | $\nu_{E3,L}$ | 523.33 | 29.59 | $\nu_{E3,L}$ | 523.33 | 75.06 |
| $\nu_{E4,H}$ | 700 | 7.39 | $\nu_{E4,H}$ | 700 | 29.59 | $\nu_{E4,H}$ | 700 | 75.06 |

Using parameters in Table 2 the normal (lattice) heat capacities were calculated, see Fig. 12. It is apparent that they almost perfectly overlap in the experimental window of 0.36 - 100 K. It is only above 110 K that they start to diverge satisfying the relation $C_{\text{normal}}(\textbf{Fe}_2\textbf{Nb}) < C_{\text{normal}}(\textbf{CoFeNb}) < C_{\text{normal}}(\textbf{Co}_2\textbf{Nb})$. The thus obtained normal heat capacities were used to extract the magnetic contributions to the heat capacity $C_{\text{mag}} = C_p - C_{\text{normal}}$. The results of the subtractions are shown in Fig. 13. In zero-field we observe clear maxima at 4.87(8) K, 7.1(2) K, and 8.44(3) K for **Co₂Nb**, **CoFeNb**, and **Fe₂Nb**, respectively, determined by the roots of d$C_{\text{mag}}$/d$T$ and corresponding to the transition points to the long-range magnetically ordered phase. Being only slightly shifted toward lower temperatures, they compare well with the values implied by the total heat capacities (see Fig. 3) and should be treated as the best estimates of the positions of the second order phase transitions observed in the studied samples. With increasing magnetic field the peaks shift toward high temperatures becoming more and more dispersed and finally they disappear in the field of 9 T. Figure 13 shows the data in the restricted range of 0-23 K, however, the zero-field data extend further up the temperature scale displaying nonzero values up to $T_f$=40.4 K, 82.8 K, and 99.0 K for **Co₂Nb**, **CoFeNb**, and **Fe₂Nb**, respectively.

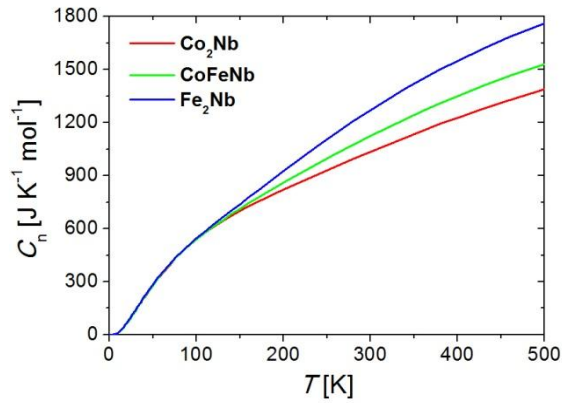

**Fig. 12:** The calculated normal heat capacities of the studied compounds.

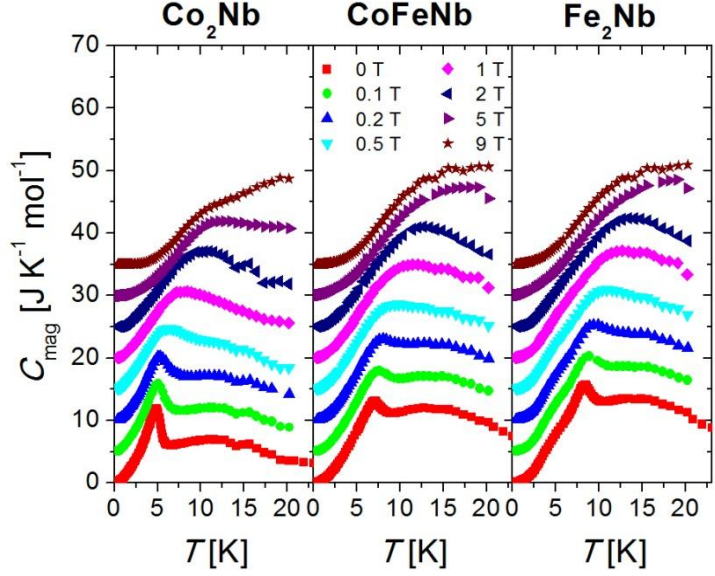

**Fig. 13:** Magnetic contributions to the heat capacity for the studied compounds. For the sake of clarity the data sets corresponding to the nonzero field values have been shifted up successively by 5 units.

To estimate the entropies associated with the zero-field magnetic heat capacities one needs to approximate the low-temperature behavior of the latter. We used the function

$$C_{mag,LT} = \frac{a}{T^2}\exp\left(-\frac{b}{T}\right) + \frac{d}{T^2},  \quad (25)$$

where the first term corresponds to the low-temperature magnetic contribution from the electronic degrees of freedom, while the second term accounts collectively for the hyperfine splitting of the $^{59}$Co and $^{93}$Nb nuclei. Fitting function in Eq. (25) to the magnetic heat capacities in the range 0.36-0.49 K one obtains $a$=4.2(4) J K mol$^{-1}$, $b$=2.50(5) K, $d$=0.0116(2) J K mol$^{-1}$, $a$=1.57(9) J K mol$^{-1}$, $b$=2.09(3) K, $d$=0.0053(2) J K mol$^{-1}$, and $a$=1.41(9) J K mol$^{-1}$, $b$=2.03(3) K, $d$=0.0038(2) J K mol$^{-1}$ for **Co$_2$Nb**, **CoFeNb**, and **Fe$_2$Nb**, respectively. The total magnetic entropy involves four contributions: the first contribution $\Delta S_1$ originates from the extrapolation of the magnetic heat capacity down to zero temperature, it is obtained by logarithmically integrating the first term in Eq. (25) with the best-fit parameter values in the range of 0-0.36 K; the second contribution $\Delta S_2$ is calculated by numerically integrating the magnetic heat capacity in the range of 0.36 K-$T_f$, $\Delta S_2 = \int_{0.36}^{T_f} C_{mag}(T)\mathrm{d}\ln T$; the third contribution $\Delta S_3$ is due to the high-temperature magnetic heat capacity tail, it is obtained by logarithmically integrating the second term in Eq. (23) in the temperature range $T_f$-∞ yielding $\Delta S_3 = cT_f^2/2$; and finally the last contribution $\Delta S_4$ originates from the high-temperature tail of the nuclear contribution to the magnetic heat capacity given by the second term in Eq. (25), it is calculated by logarithmically integrating this term with the best-fit parameter value in the range of 0.36 K-∞ giving $\Delta S_4$=3.85802$d$ and should be subtracted from the total leaving exclusively the contributions from the electronic degrees of freedom, $\Delta S_{mag}=\Delta S_1+\Delta S_2+\Delta S_3-\Delta S_4$. Table 3 collects the results of magnetic entropy calculations for the studied compounds.

**Table 3:** The results of magnetic entropy calculations; $\Delta S_{mag}=\Delta S_1+\Delta S_2+\Delta S_3-\Delta S_4$.

| entropy [J K$^{-1}$ mol$^{-1}$] | Co$_2$Nb | CoFeNb | Fe$_2$Nb |
|---|---|---|---|
| $\Delta S_1$ | 0.0052 | 0.0074 | 0.0082 |
| $\Delta S_2$ | 17.6008 | 24.7089 | 26.7973 |
| $\Delta S_3$ | 0.4607 | 0.2850 | 0.2311 |
| $\Delta S_4$ | 0.0446 | 0.0203 | 0.0147 |
| $\Delta S_{mag}$ | 18.02 | 24.98 | 27.07 |
| $\Delta S_{calc}$ | 17.29 | 24.91 | 32.53 |

The last row in Table 3 shows the expected magnetic entropies due to the electronic degrees of freedom, calculated assuming the following spin states of the constitutive ions: $S_{Co}=1/2$, $S_{Fe}=2$, and $S_{Nb}=1/2$. These values compare well with the experimental ones for **Co$_2$Nb**, **CoFeNb**, while for **Fe$_2$Nb** the experimental value is largely underestimated. This may be attributed to a possible overestimation of the normal heat capacity in this case and/or to an effective attenuation of the spin of the Fe(II) ion due to a possible magnetocrystalline anisotropy. The latter would be consistent with the magnetization data at 2 K, cf. Fig. 6(b) in [54], where the theoretical mean-field curve (black solid line) obtained assuming the high spin state of the Fe(II) ion, i.e. $S_{Fe}=2$, systematically exceeds the experimental points (blue full circles). Indeed, $\Delta S_{calc}$ calculated with $S_{Fe}\approx3/2$ amounts to 28.82 J K$^{-1}$ mol$^{-1}$, which is closer to the experimental value of 27.07 J K$^{-1}$ mol$^{-1}$. Except for **Fe$_2$Nb** the magnetic entropies inferred from the experimental data, $\Delta S_{mag}$, exceed those calculated on the basis of the presupposed spin content, $\Delta S_{calc}$. This is due to the assumption that the spin of the Co(II) ion is equal to $S_{Co}=1/2$, which corresponds to the ground Kramers doublet being exclusively populated. This is the case at sufficiently low temperatures, however, at higher temperatures the excited Kramers doublets become more and more populated enhancing the spin value toward the free-ion one $S_{Co}=3/2$.

5. Magnetocaloric effect

5.1 Isothermal entropy change $\Delta S_M$ and adiabatic temperature change $\Delta T_{ad}$

The two main MCE characteristics of the studied compound, i.e. the temperature dependence of the isothermal entropy change $\Delta S_M$ and the adiabatic temperature change $\Delta T_{ad}$, were determined indirectly using the measured heat capacity values. $\Delta S_M$ was calculated using the formula

$$\Delta S_M(T,\Delta H=0\rightarrow H)\equiv \Delta S_M(T,H)=\int_{T_{min}}^{T}\frac{C_p(T',H)-C_p(T',H=0)}{T'}dT', \quad (26)$$

where $T_{min}$ denotes the left boundary of the experimental window. In this way we deliberately neglect the entropy contribution in the interval $(0,T_{min})$ which involves both the nuclear contribution as well as the contribution from electronic degrees of freedom. While the former contribution is difficult to follow especially in nonzero magnetic field, the uncertainty associated with the latter one is on the order of the contribution $\Delta S_1$ in Table 3. The adiabatic temperature change $\Delta T_{ad}$ was estimated on the basis of the formula

$$\Delta T_{ad}(H_i \rightarrow H_f) = [T(S, H_f) - T(S, H_i)]_S, \quad (27)$$

which requires the inversion of the temperature dependence of the total entropy $S(T,H)$ calculated similarly with the lower cutoff temperature $T_{min}$. While $\Delta S_M$ was calculated in the magnetization direction, where the magnetic field change consists in switching on the field $H_i=0 \rightarrow H_f=H$, $\Delta T_{ad}$ was calculated additionally in the demagnetization direction involving switching off of the field, i.e. $H_i=H \rightarrow H_f=0$. Figure 14 collects the corresponding results.

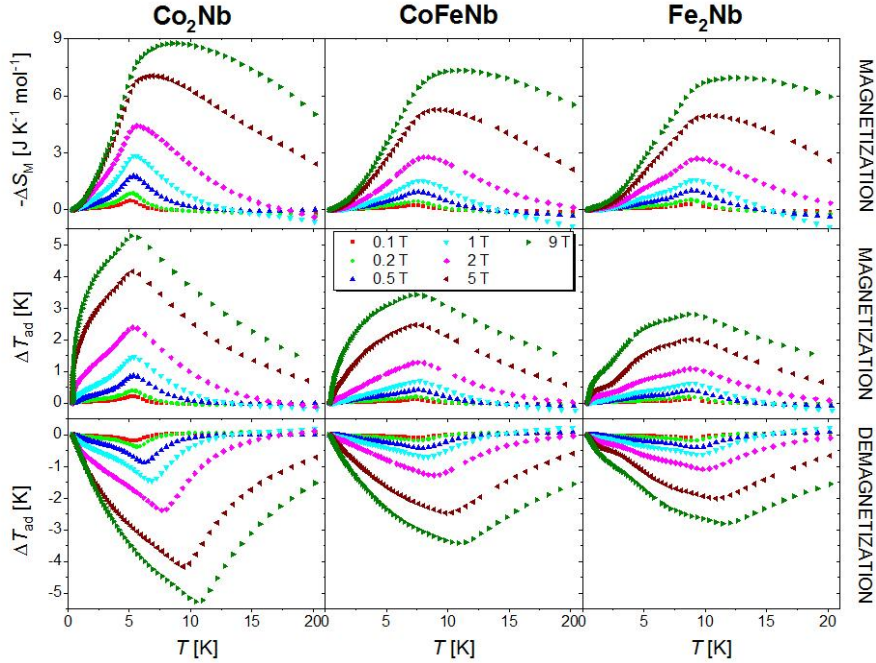

**Fig. 14:** Temperature dependence of the isothermal entropy change $\Delta S_M$ in the magnetization direction (upper panel), the adiabatic temperature change $\Delta T_{ad}$ (in the magnetization direction (middle panel), and the adiabatic temperature change $\Delta T_{ad}$ in the demagnetization direction (lower panel) for an indicated array of magnetic field change values.

It is apparent from Fig. 14 that while the amplitudes of both quantities increase with increasing field change values, they show a deceasing trend when looking from compound **Co$_2$Nb** through **CoFeNb** to **Fe$_2$Nb**. The former behavior may be attributed to the effect of quenching the magnetic entropy by the application of the magnetic field, i.e. the stronger the field the larger the quench. To understand the latter effect let us note that the signals of $|\Delta S_M|$ and $|\Delta T_{ad}|$ are consistently centered around the corresponding phase transition temperatures which shift rightward from sample to sample. In addition, Fig. 13 implies that the thermal effects $\Delta Q(H) = \int C_{mag}(H) dT$ associated with the three samples are comparable. Then the entropic effect $\Delta S(T,H) \propto \Delta Q(H)/T$ becomes roughly inversely proportional to temperature, i.e. the higher the temperature the smaller the entropy changes. Table 4 collects the peak values of the quantities $\Delta S_M$ and $\Delta T_{ad}$ for the three compounds.

**Table 4:** Peak values of quantities $\Delta S_M$ and $\Delta T_{ad}$.

| $\mu_0 \Delta H$ [T] | magnetization | | magnetization | | demagnetization | |
|---|---|---|---|---|---|---|
| | $T_{max}$ [K] | $|\Delta S_M|^{max}$ [J K$^{-1}$ mol$^{-1}$] | $T_{max}$ [K] | $\Delta T_{ad}^{max}$ [K] | $T_{max}$ [K] | $|\Delta T_{ad}|^{max}$ [K] |
| **Co$_2$Nb ($T_c$=4.87(8) K)** | | | | | | |
| 0.1 | 5.08 | 0.46 | 5.29 | 0.20 | 5.62 | 0.20 |
| 0.2 | 5.34 | 5.33 | 5.24 | 0.39 | 5.62 | 0.39 |
| 0.5 | 5.33 | 1.78 | 5.34 | 0.87 | 6.21 | 0.87 |
| 1 | 5.60 | 2.83 | 5.39 | 1.46 | 6.85 | 1.46 |
| 2 | 5.59 | 4.41 | 5.25 | 2.38 | 7.63 | 2.38 |
| 5 | 6.91 | 7.04 | 5.25 | 4.16 | 9.41 | 4.16 |
| 9 | 8.94 | 8.75 | 5.15 | 5.28 | 10.43 | 5.28 |
| **CoFeNb ($T_c$=7.1(2) K)** | | | | | | |
| 0.1 | 7.23 | 0.23 | 7.33 | 0.10 | 7.48 | 0.10 |
| 0.2 | 7.22 | 0.43 | 7.43 | 0.19 | 7.62 | 0.19 |
| 0.5 | 7.61 | 0.94 | 7.59 | 0.42 | 8.01 | 0.42 |
| 1 | 7.99 | 1.53 | 7.72 | 0.69 | 8.41 | 0.69 |
| 2 | 7.99 | 2.77 | 7.57 | 1.28 | 8.86 | 1.28 |
| 5 | 9.32 | 5.26 | 7.34 | 2.47 | 9.81 | 2.47 |
| 9 | 10.88 | 7.33 | 7.45 | 3.42 | 10.88 | 3.42 |
| **Fe$_2$Nb ($T_c$=8.44(3) K)** | | | | | | |
| 0.1 | 8.89 | 0.27 | 8.78 | 0.11 | 8.89 | 0.11 |
| 0.2 | 8.89 | 0.50 | 8.70 | 0.19 | 8.89 | 0.19 |
| 0.5 | 8.88 | 1.02 | 8.95 | 0.40 | 9.35 | 0.40 |
| 1 | 9.32 | 1.56 | 8.70 | 0.62 | 9.32 | 0.62 |
| 2 | 9.32 | 2.69 | 8.73 | 1.08 | 9.81 | 1.08 |
| 5 | 10.32 | 4.93 | 8.86 | 2.01 | 10.87 | 2.01 |
| 9 | 12.05 | 6.94 | 8.65 | 2.80 | 11.45 | 2.80 |

Table 4 shows that the peak temperatures $T_{max}$ are all placed above the transition temperature points $T_c$. For the isothermal entropy change $|\Delta S_M|$ as well as the adiabatic temperature change $|\Delta T_{ad}|$ in the demagnetization mode the peak temperatures shift visibly toward higher temperatures with the increasing field change values. In the case of $\Delta T_{ad}$ in the magnetization mode no clear trend in the dependence of $T_{max}$ on $\mu_0 \Delta H$ can be observed with the peak temperatures being stiffly anchored slightly above $T_c$. The values of $|\Delta S_M|^{max}$ detected for $\mu_0 \Delta H$=5 T amount to 7.04, 5.26, and 4.93 J K$^{-1}$ mol$^{-1}$ for **Co$_2$Nb**, **CoFeNb**, and **Fe$_2$Nb**, respectively, and are on the order of those obtained for the same field change in the isostructural compounds $\{[M^{II}(pyrazole)_4]_2[Nb^{IV}(CN)_8]\cdot 4H_2O\}_n$ with M=Mn (6.83 J K$^{-1}$ mol$^{-1}$, $T_c$=22.8 K) and M=Ni (6.1 J K$^{-1}$ mol$^{-1}$, $T_c$=13.4 K) [50, 51]. Although the spin value of the Ni(II) centre ($S_{Ni}$=1) is lower than that of the Fe(II) ion ($S_{Fe}$=2) the compound with M=Ni shows larger MCE effect. This may be due to both the local anisotropy of the Fe(II) centre, which is known to affect adversely the MCE effect [42], as well as the fact that the exchange coupling in the compound containing Ni(II) is of ferromagnetic character [54]. A similar difference occurs for the compounds with M=Co and M=Mn, where the effective spin of the Co(II) ion in the low temperature regime is as low as $S_{Co}$=1/2, while the spin of the Mn(II) ion is $S_{Mn}$=5/2, yet we observe a larger $|\Delta S_M|^{max}$ for the first sample than for the other. This may be understood by remembering that the Mn compound orders at considerably larger temperature (ca 4.7 times) than the **Co$_2$Nb**, and the entropic effect is roughly inversely proportional to temperature as discussed above. The values of $\Delta T_{ad}$ detected for $\mu_0\Delta H$=5 T amount to 4.16, 2.47, and 2.01 J K$^{-1}$ mol$^{-1}$ for **Co$_2$Nb**, **CoFeNb**, and **Fe$_2$Nb**, respectively, and are larger or comparable to those observed for the same field change for the isostructural compounds {[M$^{II}$(pyrazole)$_4$]$_2$[Nb$^{IV}$(CN)$_8$]·4H$_2$O}$_n$ with M=Mn (1.42 K) and M=Ni (2.0 K) [50, 51]. They also exceed those found for Mn$_2$-pyridazine-[Nb(CN)$_8$] (1.5 K for $\mu_0\Delta H$=5 T) [52] and hexacyanochromate Prussian blue analogues (1.2 K for $\mu_0\Delta H$=7 T) [47].

It is interesting to look at MCE for the lowest field change values $\mu_0\Delta H$=0.1, 0.2, 0.5, and 1 T, see Fig. 14. It can be seen that in these cases the inverse MCE (heating under adiabatic demagnetization) is present above the transition temperature for the three compounds. The effect increases with increasing field change value to diminish or completely vanish for $\mu_0\Delta H \geq 2$ T. A similar effect was observed for the Mn analogue [51]. A possible explanation could be that the low field is not able to reorient the correlated clusters above the transition temperature and the only effect it can cause is the local flipping of the magnetic moments, which can be seen as an disordering factor. Thus there is an additional contribution to the in-field entropy so that the entropy in zero-field is slightly smaller than that in a nonzero field.

5.2 Field dependence of MCE

The field dependence of MCE has been studied intensively either experimentally [45, 51, 70, 71] or from a theoretical viewpoint by using the description within the mean-field model [72] or employing the equation of state of materials with the second-order magnetic phase transition [73, 74]. The parameter conveniently quantifying the local sensitivity of the isothermal entropy change to the external field amplitude $H=\Delta H=H_f-H_i(=0)$ is exponent $n$ defined by the following derivative

$$n = \frac{d\ln\Delta S_M}{d\ln H}.\qquad(28)$$

The value of $n$ means that in the vicinity of a given thermodynamic point ($T,H$) the entropy change behaves approximately as $H^n$. The high-temperature limit of an $n$ vs. $T$ curve, $n$=2, is the consequence of the Curie-Weiss law, where the magnetic entropy change using the equation

$$\Delta S_M(T,\Delta H = H_f - H_i) = \int_{H_i}^{H_f} \frac{\partial M(T,H)}{\partial T}dH \qquad(29)$$

implied by the Maxwell relation, yields a quadratic field dependence of $\Delta S_M$. The value of $n$ in the low-temperature regime cannot be easily predicted, however, if the magnetization displays only a weak field dependence there, Eq. (29) implies $n$=1. The temperature dependence of the field-averaged value of exponent $n$ for the compounds under study, estimated on the basis of the entropy data in Fig. 14, is shown in Fig. 15. Parameter $n$ displays a smooth decrease on cooling from the values close to 2, in agreement with what might be expected. Next, it attains the minimum of 0.59, 0.71, and 0.68 at 5.08 K, 7.21 K, and 8.47 K, i.e. slightly above the transition temperature, for **Co$_2$Nb**, **CoFeNb**, and **Fe$_2$Nb**, respectively. In the low-temperature regime exponent $n$ assumes again a decreasing trend, displaying the second minimum at the lowest temperatures but for compound **Fe$_2$Nb**. The value of exponent $n$ at the transition temperature $T_c$ has been demonstrated to be related to the critical exponents of a material [74]:

$$n\big|_{T_c} = 1 + \frac{\beta - 1}{\beta + \gamma}. \qquad (30)$$

The values of $n$ at $T_c$ practically coincide with its values at the minima being equal to 0.59, 0.72, and 0.68 for **Co$_2$Nb**, **CoFeNb**, and **Fe$_2$Nb**, respectively. They are close to the value of 0.6424(4) obtained with Eq. (30) and the theoretical estimates for the three-dimensional (3D) Heisenberg universality class [75], namely $\beta$=0.3689(3) and $\gamma$=1.3960(9).

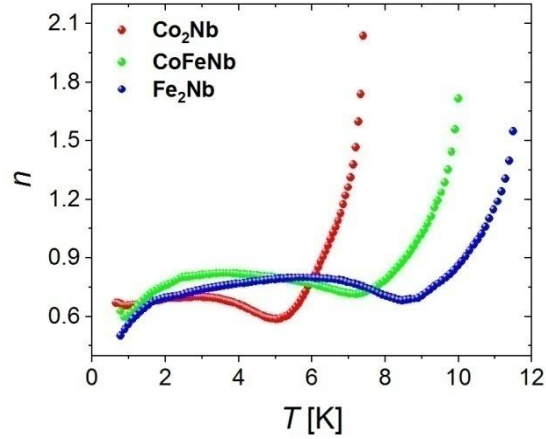

**Fig. 15:** Temperature dependence of exponent $n$ quantifying the field dependence of the isothermal entropy change $\Delta S_M$ for the three studied compounds.

5.3 Refrigeration capacity (RC)

A commonly accepted measure for the performance of a substance undergoing a magnetic cooling cycle is the refrigeration capacity (RC), defined as [76, 77]

$$\text{RC} = \int_{T_C}^{T_H} \Delta S_M(T, \Delta H) dT, \qquad (31)$$

where the temperatures $T_C$ and $T_H$ of the cold and hot reservoir, respectively, are usually selected as to cover the full-width at half-maximum of the entropy change peak $\Delta S_M(T,\Delta H)$. This quantity is a measure of how much heat can be transferred between the cold and hot reservoirs in one ideal refrigeration cycle. Table 5 collects the values of $T_C$ and $T_H$ determined in the above way for all field change values and all the studied compounds.

**Table 5:** Temperatures $T_C$ and $T_H$. The values marked by an asterisk correspond to the endpoint entropy data placed above the half-maximum level.

| $\mu_0\Delta H$ [T] | Co$_2$Nb | | CoFeNb | | Fe$_2$Nb | |
|---|---|---|---|---|---|---|
| | $T_C$ [K] | $T_H$ [K] | $T_C$ [K] | $T_H$ [K] | $T_C$ [K] | $T_H$ [K] |
| 0.1 | 4.11 | 5.97 | 5.57 | 8.55 | 7.02 | 10.01 |
| 0.2 | 4.00 | 6.30 | 5.30 | 8.91 | 6.83 | 10.37 |
| 0.5 | 3.88 | 7.20 | 5.20 | 9.91 | 6.35 | 11.33 |
| 1 | 3.69 | 8.28 | 5.08 | 10.84 | 5.84 | 12.21 |
| 2 | 3.52 | 10.29 | 5.01 | 13.27 | 5.63 | 14.85 |
| 5 | 3.62 | 17.17 | 4.98 | 18.77 | 5.71 | 20.18* |
| 9 | 3.89 | 20.16* | 5.16 | 20.23* | 5.79 | 20.21* |

In Fig. 16 the field dependence of RC is shown for the three studied samples. As might be expected RC is an increasing function of the field change value. For **CoFeNb** and **Fe$_2$Nb** the RC curves almost coincide, while for **Co$_2$Nb** the RC values exceed those calculated for the remaining compounds. RC for the field change of $\mu_0\Delta H$=5 T amounts to 72.7, 56.5, and 56.3 J kg$^{-1}$ for **Co$_2$Nb**, **CoFeNb**, and **Fe$_2$Nb**, respectively. These values are placed below those found for the same field change for the Mn (132.9 J kg$^{-1}$) and Ni (73.1 J kg$^{-1}$) analogues [50].

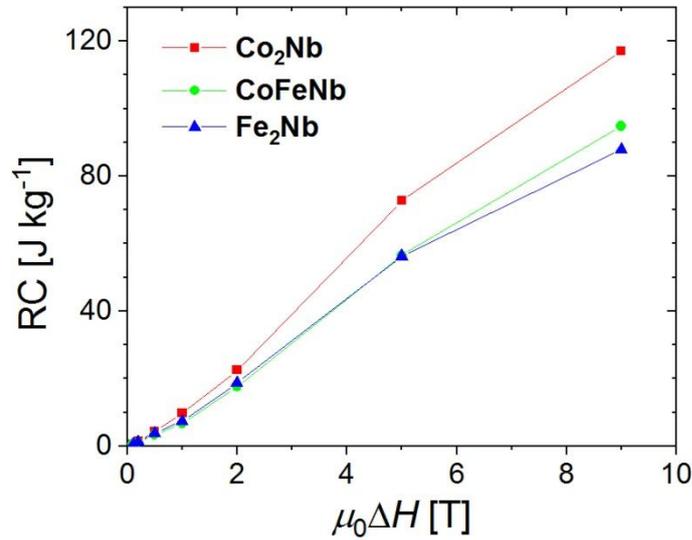

**Fig. 16:** Field dependence of the refrigerant capacity for the studied samples.

5.4 The Ericsson cycle

Among the thermodynamic cycles, important for magnetic refrigeration, the regeneration Ericsson cycle is especially worth mentioning [78, 79]. Let us consider two regeneration Ericsson cycles, referred to in what follows as Ericsson 1 and Ericsson 2 and depicted in the (*T*,*S*) plane of Fig. 17 in blue and red, respectively. Cycle Ericsson 1 (A→B→D→E→A) consists of two isothermal processes (A→B at temperature $T_H$ and D→E at temperature $T_C$) and two iso-field processes (B→D at applied field *H* and E→A at zero applied field). Similarly, cycle Ericsson 2 (F→C→D→E→F) consists of two isothermal processes (F→C at temperature $T_{max}$ and D→E at temperature $T_C$) and two iso-field processes (C→D at applied field *H* and E→F at zero applied field). Temperatures $T_C$ and $T_H$ of the cold and hot reservoirs correspond to the endpoints of the full-width at half-maximum of the entropy change peak |$\Delta S_M(T,H)$|, see Table 5, while $T_{max}$ is the peak temperature of $\Delta S_M(T,H)$, see the second column in Table 4.

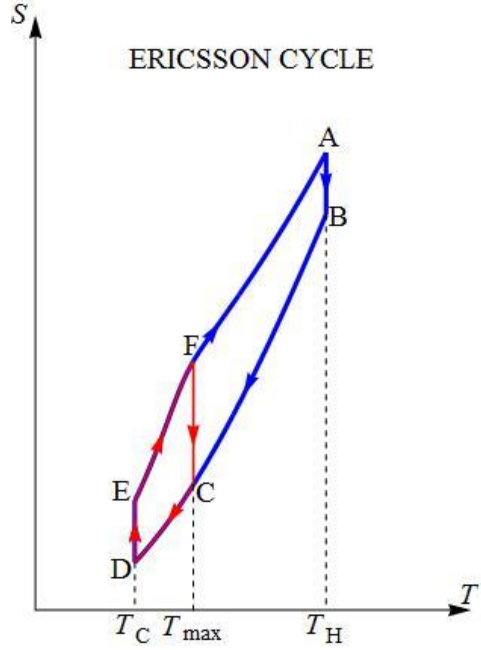

**Fig. 17:** The regeneration Ericsson cycles operating between $T_C$ and $T_H$ (Ericsson 1, blue) and $T_C$ and $T_{max}$ (Ericsson 2, red). The genuine entropy data in $\mu_0 H=0$ and 2 T for **Co$_2$Nb** have been used.

Let us first analyze cycle Ericsson 1. For this cycle the heat $Q_C$ absorbed during the isothermal process D→E and the heat $Q_H$ rejected at the isothermal process A→B can be calculated by

$$Q_C = \int_{D\to E} TdS = -T_C \Delta S_M(T_C, H) > 0,$$
$$Q_H = \int_{A\to B} TdS = T_H \Delta S_M(T_H, H) < 0, \tag{32}$$

where the isothermal entropy change $\Delta S_M(T,H)$ is given in Eq. (26). In the present case, where $T_C < T_{max} < T_H$, the non-perfect regeneration heat quantity $\Delta Q$ must be divided in two parts $\Delta Q = -(Q_{EA} + Q_{BD}) = \Delta Q^+ + \Delta Q^-$, where

$$\Delta Q^- = \int_{T_C}^{T_{max}} T \frac{\partial \Delta S_M(T,H)}{\partial T} dT < 0 \tag{33}$$

quantifies the heat that has to be compensated by the hot reservoir and

$$\Delta Q^+ = \int_{T_{max}}^{T_H} T \frac{\partial \Delta S_M(T,H)}{\partial T} dT > 0 \tag{34}$$

quantifies the heat that has to be released to the cold reservoir otherwise the temperature of the regenerator will be changed and the cycle would not operate properly. $\Delta Q^+$ corresponds to the situation where the heat transferred from the working substance to the regenerator is larger than that transferred from the regenerator to the working substance. In the case of $\Delta Q^-$ just the reverse holds, i.e. the heat transferred from the working substance to the regenerator is smaller than that transferred from the regenerator to the working substance. In this way, the net cooling quantity $Q_L$ is reduced as compared to $Q_C$, i.e. $Q_L=Q_C-\Delta Q^+$. According to the first law of thermodynamics, the work input of the refrigeration cycle W is given by the formula

$$W = -(Q_C + Q_H - \Delta Q) = T_C \Delta S_M(T_C, H) - T_H \Delta S_M(T_H, H) + \int_{T_C}^{T_H} T \frac{\partial \Delta S_M(T,H)}{\partial T} dT. \quad (35)$$

Finally, the coefficient of performance (COP) of the refrigeration cycle is given by

$$\text{COP} = \frac{Q_L}{W}. \quad (36)$$

The discussion of cycle Ericsson 2 would be fully analogous. It is sufficient to take into account that $T_H=T_{max}$ in this case. Immediately Eq. (34) implies that quantity $\Delta Q^+$ vanishes and, consequently, the net cooling quantity $Q_L=Q_C$ is not reduced. One therefore should expect enhanced COP values. The COP values for the Ericsson cycles should be compared to that of the Carnot cycle operating between the same temperatures $T_C$ and $T_H$ of the cold and hot reservoirs. The main difference is that the iso-field processes of the Ericsson cycle are replaced by the adiabatic ones. This is formally equivalent to the situation where the $\Delta S_M(T,H)$ curve is flat between $T_C$ and $T_H$, i.e. then $\partial \Delta S_M(T,H)/\partial T = 0$ and Eqs. (32)-(36) imply that

$$\text{COP}_{\text{Carnot}} = \frac{T_C}{T_H - T_C}. \quad (37)$$

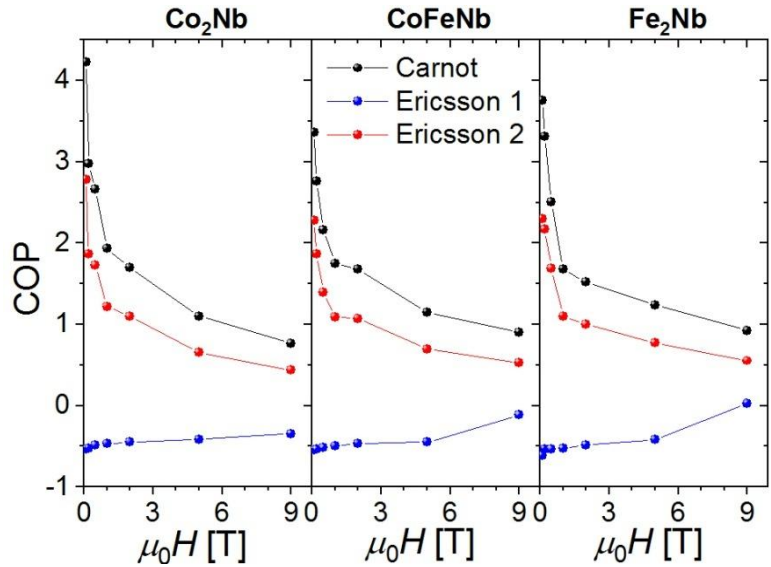

**Fig. 18:** Field dependence of the coefficient of performance of three indicated thermodynamic cycles for the studied compounds.

Figure 18 shows the field dependence of the coefficient of performance of cycles Ericsson 1, Ericsson 2, and the Carnot cycle operating between temperatures $T_C$ and $T_H$ for the studied compounds. It can be seen that COPs for all three samples display a similar pattern. Quite surprisingly, COP of Ericsson 1 assumes systematically negative values except for that for **Fe₂Nb** at $\mu_0 H=9$ T, where it turns slightly positive. This implies that this refrigeration cycle, operating in the full width of half-maximum of $|\Delta S_M(T,H)|$, becomes totally ineffective, although it improves steadily with increasing magnetic field. Shifting the temperature of the hot reservoir $T_H$ down to $T_{max}$, which is the case for Ericsson 2, remedies the situation and the corresponding COP is positive and lower than that of the Carnot cycle. However, it shows a decreasing trend with increasing magnetic field. At the same time, the refrigerant capacity RC was demonstrated to be an increasing function of the field change $\mu_0 \Delta H$, see Fig. 16, assuming very small values for the lowest field change values. In view of these facts it is clear that a compromise must be made for the Ericsson 2 cycle to be most efficient. A possible candidate quantifying its efficiency is the product RC×COP whose largest value should roughly indicate the most efficient cycle. Fig. 19 shows the field dependence of the RC×COP product for the three studied compounds. It implies that cycle Ericsson 2 should be most efficient for the maximal studied value of the applied field (=9 T) with irrelevant differences between the compounds. At the same time, the Ericsson 2 cycle using **Fe₂Nb** as the working substance at 9 T would be less efficient than that using **Co₂Nb** at the lower field value of 5 T.

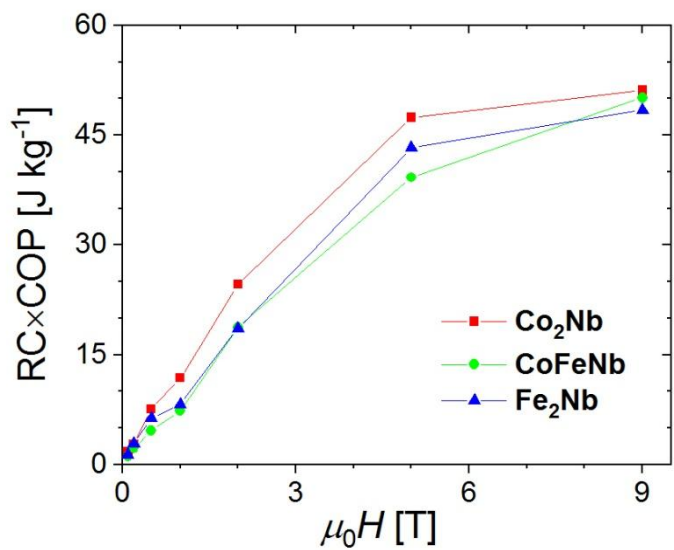

**Fig. 19:** Field dependence of product RC×COP quantifying the efficiency of cycle Ericsson 2 for the studied compounds.

6. Conclusions

We have reported a comprehensive study of thermodynamic properties of three samples of bimetallic molecular magnets $[Co^{II}(pyrazole)_4]_{2x}[Fe^{II}(pyrazole)_4]_{2(1-x)}[Nb^{IV}(CN)_8]\cdot 4H_2O$ with $x=0$, 0.5, and 1, where the middle compound represents a substitutional mixture of the two marginal ones. The three samples display the same crystallographic structure. Their heat capacities were measured in the temperature range 0.36-100 K without applied field as well as in the field of $\mu_0 H=0.1$, 0.2, 0.5, 1, 2, 5, and 9 T. The results revealed anomalies assigned to the second-order phase transitions to magnetically ordered phases with the transition temperatures estimated to amount to 4.87(8) K, 7.1(2) K, and 8.44(3) K for $x$=0, 0.5, and 1, respectively. The heat capacity results were analyzed to discuss the stability of the mixed compound and the magnetocaloric effect. The Gibbs energy of mixing turned out to be positive but smaller in magnitude than the energy of thermal fluctuations, which implies that the mixed sample is marginally stable in the full detected temperature range. There is therefore no surprise that other mixtures with $x \neq 0.5$ could not be obtained. The negative values of the enthalpy of mixing are explained in terms of an ordered arrangement of the Co and Fe ions in the solid solution **CoFeNb**. The negative values of the entropy of mixing can be rationalized by supposing that the lattice of the solid solution sample becomes more rigid due to the requirement of accommodating two different atoms within the same lattice structure. To extract the magnetic contribution to the heat capacity we took an approach whose main principle is to calculate the lattice contribution based on a reasonable frequency spectrum. While $C_{mag}$ is consistent with the expected spin content of the Co(II) and Nb(IV) ions ($S_{Co}$=1/2, $S_{Nb}$=1/2), it suggests an effective attenuation of the spin of the Fe(II) ion ($S_{Fe}$=2) due to a possible magnetocrystalline anisotropy. Taking advantage of the in-field heat capacity measurements the magnetocaloric effect was described in terms of the isothermal entropy change $\Delta S_M$ and the adiabatic temperature change $\Delta T_{ad}$. The magnitudes of these quantities turned out to be typical for the class of molecular magnets. The analysis of the field dependence of $\Delta S_M$ implied that the studied compounds belong to the universality class of the 3D Heisenberg model. To get some practical insight into the issue of magnetic refrigeration the regeneration Ericsson cycles employing the studied compounds as the working substance were considered. Most surprisingly, the Ericsson cycle operating between the temperatures corresponding to the full width at half-maximum of the $|\Delta S_M|$ signal turned out to be completely ineffective. It was only by shifting the temperature of the hot reservoir $T_H$ down to the temperature $T_{max}$ corresponding to $|\Delta S_M|^{max}$ that the coefficient of performance became positive and comparable to that of the Carnot cycle. To determine the optimal value of the magnetic field for which the Ericsson cycle is most efficient, the product of the refrigerant capacity and the coefficient of performance was employed. The calculations indicated that the regeneration Ericsson cycle operating between $T_C$ and $T_{max}$ should be most efficient for the maximal studied value of the applied field (=9 T) with irrelevant differences between the compounds. The above findings place the studied samples among possible candidates for cryogenic refrigeration.

**Appendix:** The molecular field model for an exchange coupled system with partial substitution

Let us assume that the system consists of two sublattices A and B with stoichiometric factors $v_A$ and $v_B$ respectively. Sublattice B is occupied by the B type ions with spin $S_B$ and the Landé factor $g_B$, while sites of sublattice A are randomly filled with the A$_1$ type ions ($S_{A_1}, g_{A_1}$) and the A$_2$ type ions ($S_{A_2}, g_{A_2}$) with $x$ denoting the numer concentration of the latter. Thus the system may be described by the formula $(A_{1\,1-x}A_{2\,x})_{\mu_A}B_{\mu_B}$. The Hamiltonian pertinent to the system is given by the following formula

$$\hat{H} = -J_{A_1B}\sum_{(ij)}\hat{S}_{A_1i}\cdot\hat{S}_{Bj} - J_{A_2B}\sum_{(ij)}\hat{S}_{A_2i}\cdot\hat{S}_{Bj} + \mu_B\sum_i(g_{A_1}\hat{S}_{A_1i} + g_{A_2}\hat{S}_{A_2i} + g_B\hat{S}_{Bi})\cdot\vec{H} \quad (A1)$$

where $(ij)$ denotes the summation over the pairs of the nearest neighbors, $J_{XY}$ is the superexchange coupling constant be2tween the X type ions and the Y type ions, $\mu_B$ is the Bohr magneton, and $\vec{H}$ is the external magnetic field. We neglect the superexchange coupling within the A sublattice ($J_{AA}=0$) and the B sublattice ($J_{BB}=0$), as in the case under study the corresponding bridges involve many atoms. A full solution of the model is an extremely complex task that could only be tackled by the quantum Monte Carlo methods. We therefore decide here for a simplified approach based on the molecular field approximation (MFA). In the framework of MFA the Hamiltonian in Eq. (A1) is reduced to the following form

$$\hat{H} = g_{A_1}\mu_B\hat{S}_{A_1}\cdot\vec{H}_{A_1} + g_{A_2}\mu_B\hat{S}_{A_2}\cdot\vec{H}_{A_2} + g_B\mu_B\hat{S}_B\cdot\vec{H}_B, \quad (A2)$$

where the molecular fields $\vec{H}_X$ (X=$A_1$, $A_2$, B) read

$$\begin{aligned}\vec{H}_{A_1} &= \vec{H} + \Lambda_{A_1B}\vec{M}_B \\ \vec{H}_{A_2} &= \vec{H} + \Lambda_{A_2B}\vec{M}_B \\ \vec{H}_B &= \vec{H} + \Lambda_{BA_1}\vec{M}_{A_1} + \Lambda_{BA_2}\vec{M}_{A_2}\end{aligned} \quad (A3)$$

Quantity $\vec{M}_X$ denotes the molar magnetization of sublattice X, and the molecular field constants $\Lambda_{XY}$ read

$$\Lambda_{A_1B} = \frac{J_{A_1B}Z_{A_1B}}{N_A\mu_B^2\nu_B g_{A_1}g_B}, \quad \Lambda_{BA_1} = \frac{J_{A_1B}Z_{BA_1}}{N_A\mu_B^2\nu_A(1-x)g_{A_1}g_B},$$
$$\Lambda_{A_2B} = \frac{J_{A_2B}Z_{A_2B}}{N_A\mu_B^2\nu_B g_{A_2}g_B}, \quad \Lambda_{BA_2} = \frac{J_{A_2B}Z_{BA_2}}{N_A\mu_B^2\nu_A x g_{A_2}g_B}, \quad (A4)$$

where $Z_{XY}$ denotes the number of the nearest neighbor Y type ions of the X type ion. It is clear that $Z_{A_1B} = Z_{A_2B} = Z_{AB}$, $Z_{BA_1} = (1-x)Z_{BA}$, $Z_{BA_2} = xZ_{BA}$. Moreover, the number of the coupling connections between the A and B type ions in a mole of the compound may be written either as $N_A\nu_A Z_{AB}$ or as $N_A\nu_B Z_{BA}$, hence $\nu_A Z_{AB}=\nu_B Z_{BA}$. The above relations and Eq. (A4) imply that $\Lambda_{A_1B} = \Lambda_{BA_1} \equiv \Lambda_1$ and $\Lambda_{A_2B} = \Lambda_{BA_2} \equiv \Lambda_2$, thus there only two independent molecular field constants in the model. For an arbitrary thermodynamic point $(T,H)$ the molecular field model defined in Eqs. (A2)-(A4) should be solved by an iterative numerical method. However, in the special case where the temperature is high compared to $\max(J_{A_1B}/k_B, J_{A_2B}/k_B)$ ($k_B$-the Boltzmann constant) one can venture to calculate the magnetic susceptibility $\chi$ of the system. An additional simplification stems from the fact that the model is isotropic (all exchange interactions are of the isotropic Heisenberg type) and the scalar counterpart of Eq. (A3) can be used. Then the system is in the paramagnetic state and the molar magnetizations are directly proportional to the magnetic field, i.e. $M_X=\chi_X H_X$, where

$$\chi_X = \frac{N_A \mu_B^2 \nu_X c_X g_X^2 S_X(S_X+1)}{3k_B T} \tag{A5}$$

is the paramagnetic molar susceptibility of sublattice X, and $c_X$ is the number concentration of the X type ions ($c_{A_1}=1-x, c_{A_2}=x, c_B=1$). The system of linear equations $M_X=\chi_X H_X$ (X=$A_1$, $A_2$, B) can be solved by the Cramer method for $M_X$ and the total susceptibility of the system can be calculated as $\chi=(M_{A_1}+M_{A_2}+M_B)/H$:

$$\chi = \frac{\chi_{A_1}+\chi_{A_2}+\chi_B+2\chi_{A_1}\chi_B\Lambda_1+2\chi_{A_2}\chi_B\Lambda_2-\chi_{A_1}\chi_{A_2}\chi_B(\Lambda_1-\Lambda_2)^2}{1-\chi_{A_1}\chi_B\Lambda_1^2-\chi_{A_2}\chi_B\Lambda_2^2}. \tag{A6}$$

The MFA estimate of the transition temperature $T_c$ may be found through the straightforward algebra by solving the equation $\chi^{-1}=0$ for $T$. The $x$-dependent result may be written

$$T_c(x) = \sqrt{(1-x)[T_c(0)]^2 + x[T_c(1)]^2}, \tag{A7}$$

where the extreme transition temperatures read

$$T_c(0) = \frac{N_A \mu_B^2 g_{A_1} g_B}{3k_B} \Lambda_1 \sqrt{\nu_A \nu_B S_{A_1}(S_{A_1}+1)S_B(S_B+1)},$$

$$T_c(1) = \frac{N_A \mu_B^2 g_{A_2} g_B}{3k_B} \Lambda_2 \sqrt{\nu_A \nu_B S_{A_2}(S_{A_2}+1)S_B(S_B+1)}. \tag{A8}$$

In the case under study sublattice A corresponds to the Co sublattice in **Co₂Nb**, next it is substituted by the Fe ions, so that $A_1$=Co and $A_2$=Fe. The B sublattice corresponds to the untouched Nb ions, i.e. B=Nb. The pertinent values of parameters are: $\nu_{Co}$=2, $\nu_{Nb}$=1, $Z_{NbCo}$=4, $Z_{CoNb}$=2, $S_{Co}$=1/2, $S_{Fe}$=2, $S_{Nb}$=1/2, $g_{Co}$=5.03, $g_{Fe}$=2.16, $g_{Nb}$=2 [54].